\documentclass[twocolumn,showpacs,prl]{revtex4}

\input{epsf}

\usepackage{graphics}
\usepackage{subfigure}
\usepackage{subfigure}
\usepackage{setspace}
\usepackage{Tabbing}
\usepackage{extramarks}
\usepackage{chngpage}
\usepackage[usenames,dvipsnames]{color}
\usepackage{graphicx,float,wrapfig}
\usepackage[colorlinks,hyperindex]{hyperref}
\hypersetup{citecolor=Magenta, linkcolor=Blue, urlcolor=Red}
 \usepackage[usenames,dvipsnames]{pstricks}
 \usepackage{epsfig}
 \usepackage{pst-grad} 
 \usepackage{pst-plot} 
\begin{document}


\title{An integral formulation of Yang-Mills on loop space}

\author{L. A. Ferreira and G. Luchini}
\affiliation{Instituto de F\'\i sica de S\~ao Carlos; IFSC/USP; 
Universidade de S\~ao Paulo  \\ 
Caixa Postal 369, CEP 13560-970, S\~ao Carlos-SP, Brazil}


\begin{abstract}
It is proposed an integral formulation of classical Yang-Mills equations in the presence of sources, based on concepts in loop spaces and on  a generalization of the non-abelian Stokes theorem for two-form connections.  The formulation leads in a quite direct way to the construction of gauge invariant conserved quantities which are also independent of the  parameterization  of surfaces and volumes. Our  results  are important in understanding  global properties of non-abelian gauge theories. 
\end{abstract}


\maketitle

\def\rf#1{(\ref{eq:#1})}
\def\lab#1{\label{eq:#1}}
\def\nonu{\nonumber}
\def\br{\begin{eqnarray}}
\def\er{\end{eqnarray}}
\def\be{\begin{equation}}
\def\ee{\end{equation}}
\def\({\left(}
\def\){\right)}
\def\pa{\partial}
\def\rlx{\relax\leavevmode}
\def\IR{\rlx\hbox{\rm I\kern-.18em R}}
\def\vp{\varphi}
\def\ve{\varepsilon}

\newcommand{\sbr}[2]{\left\lbrack\,{#1}\, ,\,{#2}\,\right\rbrack}
\newcommand{\pbr}[2]{\{\,{#1}\, ,\,{#2}\,\}}
\newcommand{\pbbr}[2]{\lcurl\,{#1}\, ,\,{#2}\,\rcurl}
\newcommand\bra[1]{\langle \, {#1}\, \mid}
\newcommand\ket[1]{\mid \, {#1} \, \rangle}
\newcommand\braket[2]{\langle \, {#1}\, \mid \, {#2} \, \rangle}

\def\IZ{\rlx\hbox{\sf Z\kern-.4em Z}}
\def\IR{\rlx\hbox{\rm I\kern-.18em R}}
\def\IC{\rlx\hbox{\,$\inbar\kern-.3em{\rm C}$}}
\def\one{\hbox{{1}\kern-.25em\hbox{l}}}

The aim of the present paper is to propose an integral formulation of the classical equations of motion of non-abelian gauge theories. Our approach is based on a generalization of the non-abelian Stokes theorem  for two-form connections, which allows to present the Yang-Mills equations as the equality of an ordered volume integral to an ordered surface integral on its border. The formulation leads in a quite simple way to the construction of gauge invariant conserved quantities which are independent of the parameterizations of  volumes and surfaces.   The most appropriate mathematical language to phrase our results is that of generalized loop spaces. There is a quite vast literature on integral and loop space formulations of gauge theories \cite{loopgauge}. Our approach differs in many aspects of those formulations even though it shares some of the ideas and insights permeating them. We make however concrete progress in relation to those approaches. 
The main statement of this paper is: 

{\em Consider a Yang-Mills theory  for a gauge group $G$, with gauge field $A_{\mu}$, in the presence of matter currents $J^{\mu}$, on a  four dimensional space-time $M$. Let $\Omega$ be any tridimensional (topologically trivial)  volume on $M$, and $\partial \Omega$ be its border. We choose a reference point $x_R$ on $\partial \Omega$ and scan $\Omega$ with closed surfaces, based on $x_R$,  labelled by $\zeta$, and we scan the closed surfaces with closed loops based on $x_R$, labelled by $\tau$, and parametrized by $\sigma$, as we describe below. The classical dynamics of the gauge fields is governed by the following integral equations, on any such volume $\Omega$, 
\be
P_2 e^{ie\int_{\partial\Omega}d\tau d\sigma \left[ \alpha F_{\mu\nu}^W+\beta {\widetilde F}_{\mu\nu}^W\right] \frac{dx^{\mu}}{d\sigma}\frac{dx^{\nu}}{d\tau}}=  P_3e^{\int_{\Omega} d\zeta d\tau  V{\cal J}V^{-1}}
\lab{basic}
\ee
where $P_2$ and $P_3$ means surface and volume ordered integration respectively, ${\widetilde F}_{\mu\nu}$ is the Hodge dual of the field tensor, i.e. $F_{\mu\nu}\equiv \frac{1}{2}\,\varepsilon_{\mu\nu\rho\lambda}\, {\widetilde F}^{\rho\lambda}$,  with  $F_{\mu\nu}=\partial_{\mu}A_{\nu}-\partial_{\nu}A_{\mu}+i\,e\,\sbr{A_{\mu}}{A_{\nu}}$, $e$ is the  gauge coupling constant, $\alpha$ and $\beta$ are free parameters,  and where we have used the notation $X^W\equiv W^{-1}\,X\,W$, with $W$ being the  Wilson line defined on a curve $\Gamma$, parameterized by $\sigma$, through the equation
\be
\frac{d\,W}{d\,\sigma}+   i\,e\,A_{\mu}\,\frac{d\,x^{\mu}}{d\,\sigma}\,W=0
\lab{eqforw}
\ee
where $x^{\mu}$ ($\mu=0,1,2,3$) are the coordinates on the  four dimensional space-time $M$. The quantity $V$ is defined on a surface $\Sigma$ through the equation
\be
\frac{d\,V}{d\,\tau}-V\, T\(A,\tau\)=0
\lab{eqforv}
\ee
with $ T\(A,\tau\)\equiv ie\,
 \int_{0}^{2\pi}d\sigma W^{-1}\left[ \alpha F_{\mu\nu}+\beta{\widetilde F}_{\mu\nu}\right] W \frac{dx^{\mu}}{d\sigma}\frac{dx^{\nu}}{d\tau}$. and where 
\br
&&{\cal J}\equiv
\int_0^{2\pi}d\sigma\left\{ ie\beta {\widetilde J}_{\mu\nu\lambda}^W
\frac{dx^{\mu}}{d\sigma}\frac{dx^{\nu}}{d\tau}
\frac{dx^{\lambda}}{d\zeta} +e^2\int_0^{\sigma}d\sigma^{\prime}\right. 
\lab{caljdef}\\
&&\times\left.
\sbr{\(\(\alpha-1\) F_{\kappa\rho}^W+\beta {\widetilde F}_{\kappa\rho}^W\)\(\sigma^{\prime}\)}
{\(\alpha F_{\mu\nu}^W+\beta {\widetilde F}_{\mu\nu}^W\)\(\sigma\)}\right.\nonumber\\
&&\left. \times
\, \frac{d\,x^{\kappa}}{d\,\sigma^{\prime}}\frac{d\,x^{\mu}}{d\,\sigma}
\(\frac{d\,x^{\rho}\(\sigma^{\prime}\)}{d\,\tau}\frac{d\,x^{\nu}\(\sigma\)}{d\,\zeta}
-\frac{d\,x^{\rho}\(\sigma^{\prime}\)}{d\,\zeta}\frac{d\,x^{\nu}\(\sigma\)}{d\,\tau}\)\right\}
\nonumber
\er  
where ${\widetilde J}_{\mu\nu\lambda}$ is the Hodge dual of the current, i.e. $J^{\mu}=\frac{1}{3!}\varepsilon^{\mu\nu\rho\lambda}\,{\widetilde J}_{\nu\rho\lambda}$. The Yang-Mills equations are recovered from \rf{basic} in the case where $\Omega$ is taken to be an infinitesimal volume. Under appropriate boundary conditions the conserved charges are the eigenvalues of the operator
\be
Q_S=P_2 e^{ie\int_{\partial S}d\tau d\sigma (\alpha F_{\mu\nu}^W+\beta {\widetilde F}_{\mu\nu}^W) \frac{dx^{\mu}}{d\sigma}\frac{dx^{\nu}}{d\tau}}=  P_3e^{\int_{S} d\zeta d\tau  V{\cal J}V^{-1}}
\lab{charge}
\ee
where $S$ is the $3$-dimensional spatial sub-manifold of $M$. Equivalently the charges are ${\rm Tr} Q_S^N$.  
}

In order to prove that \rf{basic} does correspond to and integral formulation of the classical Yang-Mills dynamics, we shall start by describing the generalization of the non-abelian Stokes theorem  as  formulated in \cite{afs1,afs2}. Consider a surface $\Sigma$ scanned by a set of closed loops with common base point $x_{R}$ on the border $\partial \Sigma$. The points on the loops are parameterized by  $\sigma \in [0,2\pi]$ and each loop is labeled by a parameter $\tau$ such that $\tau=0$ corresponds to the infinitesimal loop around $x_R$, and $\tau=2\pi$ to the border $\partial \Sigma$. We then introduce, on each point of $M$,  a rank two antisymmetric tensor $B_{\mu\nu}$ taking values on the Lie algebra ${\cal G}$ of $G$, and construct a quantity $V$ on the surface $\Sigma$ through   \rf{eqforv}, but with $T\(A,\tau\)$ replaced by  $T\(B,A,\tau\)\equiv
 \int_{0}^{2\,\pi}d\sigma\; W^{-1}\,B_{\mu\nu}\,W\, \frac{d\,x^{\mu}}{d\,\sigma}\,\frac{d\,x^{\nu}}{d\,\tau}$, 
and where the $\sigma$-integration is along the loop $\Gamma$ labeled by $\tau$, and $W$ is obtained from \rf{eqforw}, by integrating it along $\Gamma$ from the reference point $x_R$ to the point labeled by $\sigma$, where $B_{\mu\nu}$ is evaluated. By integrating \rf{eqforv}, from the infinitesimal loop around $x_R$ to the border of $\Sigma$, we obtain 
$V= V_R\; P_2 e^{\int_0^{2\,\pi}d\tau\int_0^{2\,\pi}d\sigma W^{-1} B_{\mu\nu}W \frac{dx^{\mu}}{d\sigma}\,\frac{dx^{\nu}}{d \tau}}$, 
where  $P_2$ means surface ordering according to the parameterization of $\Sigma$ as described above, and $V_R$ is an integration constant corresponding to the value of $V$ on an infinitesimal surface around $x_R$.
If one changes $\Sigma$, keeping its border fixed, by making variations $\delta x^{\mu}$ perpendicular to $\Sigma$ then $V$ varies according to (see sec. 5.3 of \cite{afs1},  sec. 2.3 of \cite{afs2}, or the appendix of \cite{second})
\br
&&\delta V\,V^{-1}\equiv
\int_0^{2\,\pi}d\tau\,\int_0^{2\,\pi}d\sigma\,V\(\tau\)\,\left\{
\right.\lab{deltav}\\
&& \left. W^{-1}\,
\left[D_{\lambda}B_{\mu\nu}+D_{\mu}B_{\nu\lambda}+D_{\nu}B_{\lambda\mu}\right]
\,
W\frac{d\,x^{\mu}}{d\,\sigma}\,\frac{d\,x^{\nu}}{d\,\tau}\,
\delta x^{\lambda}\right. \nonumber\\
&&\left. -\int_0^{\sigma}d\sigma^{\prime}
\sbr{B_{\kappa\rho}^W\(\sigma^{\prime}\)-ie F_{\kappa\rho}^W\(\sigma^{\prime}\)}
{B_{\mu\nu}^W\(\sigma\)}\frac{dx^{\kappa}}{d\sigma^{\prime}}\frac{dx^{\mu}}{d\sigma}\right.\nonumber\\
&&\left.\times
\(\frac{d\,x^{\rho}\(\sigma^{\prime}\)}{d\,\tau}\delta x^{\nu}\(\sigma\)
-\delta x^{\rho}\(\sigma^{\prime}\)\,\frac{d\,x^{\nu}\(\sigma\)}{d\,\tau}\)\right\} V^{-1}\(\tau\)
\nonumber
\er  
where  $D_{\mu}*=\partial_{\mu}*+i\,e\,\sbr{A_{\mu}}{*}$. The quantity $V\(\tau\)$ appearing on the r.h.s. of \rf{deltav} is obtained by integrating \rf{eqforv} from the infinitesimal loop around $x_R$ to the the loop labelled by $\tau$ on the scanning of $\Sigma$ described above. Note that the two $\sigma$-integrations on the second term on the r.h.s. of \rf{deltav} are performed on the same loop labelled by $\tau$. Consider now the case where the surface $\Sigma$ is closed, and the border of $\Sigma$ is  contracted to $x_R$. The expression \rf{deltav} gives then the variation of $V$ when we vary $\Sigma$ keeping  $x_R$ fixed. Therefore, if one starts with an infinitesimal closed surface $\Sigma_{R}$ around $x_R$ one can blows it up until it becomes $\Sigma$. One can label all those closed surfaces using a parameter $\zeta \in [0,2\pi]$, such that $\zeta=0$ corresponds to $\Sigma_{R}$ and $\zeta=2\,\pi$ to $\Sigma$. The expression \rf{deltav} can be seen as a differential equation on $\zeta$ defining $V$ on the surface $\Sigma$, i.e.
\be
\frac{d\,V}{d\,\zeta} - {\cal K}\, V=0
\lab{eqforv2}
\ee
where ${\cal K}$ corresponds to the r.h.s. of  \rf{deltav} with $\delta x^{\mu}$ replaced by $\frac{d\,x^{\mu}}{d\,\zeta}$. By integrating \rf{eqforv2} from $\Sigma_{R}$  to $\Sigma$, one obtains $V$ evaluated on $\Sigma$, which is now an ordered volume integral, over the volume $\Omega$ inside $\Sigma$, and the ordering is determined by the scanning of $\Omega$ by closed surfaces as described above. But this result has of course to be the same as that obtained by integrating \rf{eqforv} when the surface  is closed, namely $\partial\Omega$.  Therefore, we obtain the generalized non-abelian Stokes theorem for a two-form connection $B_{\mu\nu}$, parallel transported by a one-form connection $A_{\mu}$  
\be
V_R\, P_2 e^{\int_{\partial\Omega}d\tau d\sigma W^{-1}B_{\mu\nu}W \frac{dx^{\mu}}{d\sigma}\,\frac{d\,x^{\nu}}{d\,\tau}}= P_3e^{\int_{\Omega} d\zeta \,{\cal K}}\,V_{R}
\lab{stokes}
\ee
where $P_3$ means volume ordering according to the scanning described above, and $V_R$ is  the integration constant obtained when integrating \rf{eqforv} and \rf{eqforv2}. It corresponds in fact to the value of $V$ at the reference point $x_R$.   Note that such theorem holds true on a space-time of any dimension, and since the calculations leading to it make no mention to a metric tensor, it is valid on flat or curved space-time. The only restrictions appear when the topology of the space-time is non-trivial (existence of handles or holes for instance).  

Going back to \rf{basic} one notes that it can be obtained from \rf{stokes} by replacing  $B_{\mu\nu}$  by $ie\left[ \alpha\, F_{\mu\nu}+\beta\, {\widetilde F}_{\mu\nu}\right]$, and using  the Yang-Mills equations,
$D_{\nu}F^{\nu\mu}= J^{\mu}$ and  $D_{\nu}{\widetilde F}^{\nu\mu}=0$,  
to replace $\(D_{\lambda}B_{\mu\nu}+D_{\mu}B_{\nu\lambda}+D_{\nu}B_{\lambda\mu}\)$ in \rf{deltav} by $(-ie\beta {\widetilde J}_{\mu\nu\lambda})$, and so ${\cal K}$ introduced in \rf{eqforv2} is now given by
${\cal K}=\int_0^{2\,\pi}d\tau\, V\, {\cal J}\,V^{-1}$, 
with ${\cal J}$ given in \rf{caljdef}.
 Therefore, \rf{basic} is a direct consequence of the Yang-Mills equations  and the Stokes theorem \rf{stokes}. Note that  $V_R$  introduced in \rf{stokes}, does not appear in \rf{basic} because it has to lie in the  centre $Z\(G\)$ of $G$ to keep the gauge covariance of \rf{basic} (see \cite{gauge}). On the other hand the integral equation \rf{basic} implies the local Yang-Mills equations. In order to see that, consider the case where $\Omega$ is a infinitesimal volume of rectangular shape with lengths $dx^{\mu}$, $dx^{\nu}$ and $dx^{\lambda}$ along three chosen Cartesian axis labelled by $\mu$, $\nu$ and $\lambda$.  We choose the reference point $x_R$ to be at a vertex of $\Omega$. By considering only the lowest order contributions, in the lengths of $\Omega$, to the integrals in \rf{basic}, one observes that the surface and volume ordering become irrelevant. We have to pay attention only to the orientation of the derivatives of the coordinates w.r.t. the parameters $\sigma$, $\tau$ and $\zeta$, determined by the scanning of $\Omega$ described above. In addition, the contribution of a given face of $\Omega$ for the l.h.s. of \rf{basic} can be obtained by evaluating the integrand on any given point of the face since the differences will be of higher order.  Consider the two faces parallel to the plane $x^{\mu}x^{\nu}$. The contribution to the l.h.s. of \rf{basic} of the face at $x_R$ is   given by $-ie(\alpha F_{\mu\nu}+\beta{\widetilde F}_{\mu\nu})_{x_R}dx^{\mu}dx^{\nu}$, with the minus sign due to the orientation of the derivatives, and the contribution of the face at $x_R+dx^{\lambda}$ is 
$ie(W^{-1}(\alpha F_{\mu\nu}+\beta{\widetilde F}_{\mu\nu})W)_{(x_R+dx^{\lambda})}dx^{\mu}dx^{\nu}$, with $W_{(x_R+dx^{\lambda})}\sim \one-ieA_{\lambda}\(x_R\)dx^{\lambda}$. By Taylor expanding the second term, the joint contribution is 
$ieD_{\lambda}(\alpha F_{\mu\nu}+\beta{\widetilde F}_{\mu\nu})_{x_R}dx^{\mu}dx^{\nu}dx^{\lambda}$, with no sums in the Lorentz indices. The contributions of the other two pairs of faces are similar, and the l.h.s. of \rf{basic} to lowest order is $\one +ie(D_{\lambda}[\alpha F_{\mu\nu}+\beta{\widetilde F}_{\mu\nu}]+\mbox{cyclic perm.})_{x_R}dx^{\mu}dx^{\nu}dx^{\lambda}$.  When evaluating the r.h.s. of \rf{basic} we can take the integrand at any point of $\Omega$ since the differences are of higher order. In addition, the commutator term in ${\cal J}$ given in \rf{caljdef} is of higher order w.r.t. the first term involving the current. Therefore, the r.h.s. of \rf{basic} to lowest order is $\one+ie\beta{\widetilde J}_{\mu\nu\lambda}dx^{\mu}dx^{\nu}dx^{\lambda}$. Equating the coefficients of $\alpha$ and $\beta$ one gets  the pair of (Hodge dual) Yang-Mills equations. 

Let us discuss some consequences of \rf{basic}. In order to write it for a given volume $\Omega$,  we had to choose a reference point $x_R$ on its border, and define a scanning of $\Omega$ with surfaces and loops. If one changes the reference point and the scanning, both sides of \rf{basic} will change. However, the generalized non-abelian Stokes theorem \rf{stokes} guarantees that the changes are such that both sides are still equal to each other. Therefore, one can say that \rf{basic} transforms ``covariantly'' under the change of scanning and reference point.  In fact to be precise, the equation \rf{basic} is formulated not on $\Omega$ but on the generalized loop space $L\Omega= \left\{ \gamma: S^2 \rightarrow \Omega\, |\, {\mbox{\rm north pole}} \rightarrow x_R\in \partial\Omega\right\}$. The image of a given $\gamma$ is a closed surface $\Sigma$ in $\Omega$ containing $x_R$. A scanning of $\Omega$ is a collection of surfaces $\Sigma$, parametrized by $\tau$, such that $\tau=0$ corresponds to the infinitesimal surface around $x_R$ and $\tau =2\pi$ to $\partial\Omega$. Such collection of surfaces is a path in $L\Omega$ and each one corresponds to $\Omega$ itself. In order to perform each mapping $\gamma$ we scan the corresponding surface $\Sigma$ with closed loops starting and ending at $x_R$, and each loop is parametrized by $\sigma$, in the same way as we did in the arguments leading to \rf{stokes}. Therefore, the change of the scanning of $\Omega$ corresponds to a change of path in $L\Omega$. In this sense, the r.h.s. of \rf{basic} is a path dependent quantity in $L\Omega$ and its l.h.s. is evaluated at the end of the path. Of course, we do not want  physical quantities to depend upon the choice of paths in $L\Omega$, neither on the reference point. Note that if we take, in the four dimensional space-time $M$, a closed tridimensional volume $\Omega_c$, then the integral Yang-Mills equation \rf{basic} implies that 
\be
P_3e^{\oint_{\Omega_c} d\zeta d\tau  V{\cal J}V^{-1}}=\one
\lab{niceclosed}
\ee
since the border $\partial\Omega_c$ vanishes, and the ordered integral of the l.h.s. of \rf{basic} becomes trivial. On the loop space $L\Omega_c$, $\Omega_c$ corresponds to a closed path starting and ending at $x_R$. Consider now a point $\gamma$ on that closed path, corresponding to a closed surface $\Sigma$, in such a way that $\Omega_1$ corresponds to the first part of the path and $\Omega_2$  to the second, i.e. $\Omega_c= \Omega_1+ \Omega_2$, and $\Sigma$ is the common border of $\Omega_1$ and $\Omega_2$. By the ordering of the integration determined by  \rf{eqforv2} one observes that the relation \rf{niceclosed} can be split as $P_3e^{\int_{\Omega_2} d\zeta d\tau  V{\cal J}V^{-1}}P_3e^{\int_{\Omega_1} d\zeta d\tau  V{\cal J}V^{-1}}=\one$. However, by reverting the sense of integration along the path,  one gets the inverse operator when integrating \rf{eqforv2}. Therefore, $\Omega_1$ and $\Omega_2^{-1}$ are two different paths (volumes) joining the same points, namely  the infinitesimal surface around $x_R$ and the surface $\Sigma$, which correspond to their border. One then concludes that the operator $P_3e^{\int_{\Omega} d\zeta d\tau  V{\cal J}V^{-1}}$ is independent of the path, and so of the scanning of $\Omega$, as long as the end points, i.e. $x_R$ and the border $\partial\Omega$, are kept fixed.  

The path independency of that operator can be used to construct conserved charges using the ideas of \cite{afs1,afs2}. First of all, let us assume that the space-time is of the form ${\cal S}\times \IR$, with $\IR$ being time and ${\cal S}$ the spatial sub-manifold which we assume simply connected and without border. An example is when ${\cal S}$ is the three dimensional sphere $S^3$.  It follows from \rf{niceclosed} that $Q_{{\cal S}}\equiv P_3e^{\oint_{{\cal S}} d\zeta d\tau  V{\cal J}V^{-1}}=\one$. That means that $Q_{{\cal S}}$ is not only  conserved in time,  but also that there can be no net charge  in ${\cal S}$. In fact, there is the possibility of getting charge quantization conditions in such case  (see \cite{quantize,alvarez-olive}).  

Let us now assume the space-time is not bounded, but still simply connected, like $\IR^4$. We shall consider two paths (volumes) joining the same two points, namely the infinitesimal surface around  $x_R$, which we take to be at the time $x^0=0$, and the two-sphere at spatial infinity $S^{2,(t)}_{\infty}$, at $x^0=t$. The first path is made of two parts. The first part corresponding to the whole space at  $x^0=0$, i.e. the volume $\Omega_{\infty}^{(0)}$ inside $S^{2,(0)}_{\infty}$, the two-sphere at spatial infinity  at $x^0=0$. The second part is a hyper-cylinder $S^2_{\infty}\times I$, where $I$ is the time interval between $x^0=0$ and $x^0=t$, and $S^2_{\infty}$ is a two-sphere at spatial infinity at the times on that interval. The second path is also made of two parts.  The first one corresponds to the infinitesimal hyper-cylinder $S^2_0\times I$, where $S^2_0$ is the infinitesimal two-sphere around $x_R$ and $I$ as before.  The second part  corresponds to $\Omega_{\infty}^{(t)}$, the whole space at time $x^0=t$, i.e. the volume inside $S^{2,(t)}_{\infty}$.  From the path independency following from \rf{niceclosed} one has that the integration of \rf{eqforv2}  along those two paths should give the same result, i.e.  
$V(S^2_{\infty}\times I)\,V(\Omega_{\infty}^{(0)})=V(\Omega_{\infty}^{(t)})V(S^2_0\times I)$, 
where we have used the notation $V\(\Omega\)\equiv P_3e^{\int_{\Omega} d\zeta d\tau  V{\cal J}V^{-1}}$, and where all integrations start at the reference point $x_R$ taken to be at $x^0=0$, and at the border $S^{2,(0)}_{\infty}$ of $\Omega_{\infty}^{(0)}$. In fact, one obtains $V\(\Omega\)$ by integrating \rf{eqforv2}, and so one has to calculate  ${\cal K}=\int_0^{2\,\pi}d\tau\, V\, {\cal J}\,V^{-1}$, on the surfaces scanning the volume $\Omega$. We shall scan a hyper-cylinder $S^2\times I$ with surfaces, based at $x_R$, of the form given in figure (1.b), with $t^{\prime}$ denoting a time in the interval $I$. Each one of such surfaces are scanned with loops, labelled by $\tau$, in the following way. For $0\leq \tau\leq \frac{2\pi}{3}$, we scan the infinitesimal cylinder as shown in figure (1.a), then for $\frac{2\pi}{3}\leq \tau\leq \frac{4\pi}{3}$ we scan the sphere $S^2$ as shown in figure (1.b), and finally for $\frac{4\pi}{3}\leq \tau\leq 2\pi$ we go back to  $x_R$ with loops as shown in figure (1.c). The quantity ${\cal K}$ can then be split into the contributions coming from each one of those surfaces as ${\cal K}={\cal K}_a+{\cal K}_b+{\cal K}_c$. In the case of the infinitesimal hyper-cylinder $S^2_0\times I$, the sphere has infinitesimal radius and so it does not really contribute to ${\cal K}_b$. We shall assume the currents and field strength vanish at spatial infinity no slower than  $J_{\mu} \sim 1/R^{2+\delta}$, and $F_{\mu\nu}\sim 1/R^{\frac{3}{2}+\delta^{\prime}}$, with $\delta,\delta^{\prime}>0$, for $R\rightarrow \infty$. Therefore the quantity ${\cal J}$, given in \rf{caljdef}, vanishes when calculated on loops at spatial infinity. Consequently, in the case of the hyper-cylinder $S^2_{\infty}\times I$, the contribution to ${\cal K}_b$ coming from the sphere with infinite radius vanishes, and we have that ${\cal K}$ calculated on the surfaces scanning $S^2_{\infty}\times I$ and $S^2_0\times I$ is the same, and so $V\(S^2_{\infty}\times I\)=V\(S^2_0\times I\)$. In fact there is more to it, since when we contract the radius of the cylinders in figure 1 to zero the loops in figures (1.a) and (1.c) become the same. Therefore, the quantities ${\cal J}$ calculated on them are the same except for a minus sign coming from the derivatives $\frac{dx^{\mu}}{d\tau}$, since the loops in figure (1.a) get longer  with the increase of $\tau$, and in figure (1.c) the opposite occurs. In addition, the quantity $V$ inside the the expression  ${\cal K}=\int_0^{2\,\pi}d\tau\, V\, {\cal J}\,V^{-1}$ is insensitive to that sign since it is obtained by integrating \rf{eqforv} starting at $x_R$ in both cases. Therefore, it turns out that ${\cal K}_a+{\cal K}_c=0$. The loops scanning the sphere in figure (1.b) have legs linking the reference point $x_R$, at $x^0=0$, to the same space point but at $x^0=t^{\prime}$, i.e. $x_R^{t^{\prime}}$. Therefore, when integrating \rf{eqforv} one gets $V_{x_R}= W(x_R^{t^{\prime}},x_R)^{-1}V_{x_R^{t^{\prime}}}W(x_R^{t^{\prime}},x_R)$, where $W(x_R^{t^{\prime}},x_R)$ is obtained by integrating \rf{eqforw} along the leg linking $x_R$ to $x_R^{t^{\prime}}$, and where we have used the notation $V_x$, meaning $V$ obtained from \rf{eqforv} with reference point $x$. Using the same arguments and notation one obtains from \rf{caljdef} that, on the loops of figure (1.b), ${\cal J}_{x_R}= W(x_R^{t^{\prime}},x_R)^{-1}{\cal J}_{x_R^{t^{\prime}}}W(x_R^{t^{\prime}},x_R)$, and so  ${\cal K}_{b,x_R}= W(x_R^{t^{\prime}},x_R)^{-1}{\cal K}_{b,x_R^{t^{\prime}}}W(x_R^{t^{\prime}},x_R)$. The quantity $V(\Omega_{\infty}^{(t)})$ is obtained by integrating \rf{eqforv2} and by scanning the volume $\Omega_{\infty}^{(t)}$ with surfaces of the type shown in figure (1.b), and where the radius of $S^2$ varies from zero to infinity keeping the point $x_R^t$ fixed. Therefore, from the above arguments one gets that $V_{x_R}(\Omega_{\infty}^{(t)})=W(x_R^{t},x_R)^{-1}V_{x_R^{t}}(\Omega_{\infty}^{(t)})W(x_R^{t},x_R)$. One then concludes that such operator has an iso-spectral time evolution 
$V_{x_R^{t}}(\Omega_{\infty}^{(t)})=U(t)V_{x_R}(\Omega_{\infty}^{(0)})U(t)^{-1}$, 
 with $U(t)=W(x_R^{t},x_R)V\(S^2_{0}\times I\)$. Therefore, its eigenvalues, or equivalently ${\rm Tr}(V_{x_R^{t}}(\Omega_{\infty}^{(t)}))^N$, are constant in time. Note that from the Yang-Mills equations \rf{basic} one has that such operator can be written either as a volume or surface ordered integrals, and so we have proved \rf{charge}. We have shown that, as a consequence of \rf{niceclosed}, such operators are independent of the scanning of the volume. The reference point $x_R^t$ is on the border of the volume and so at spatial infinity.  Then when we change the reference point on the border to ${\widetilde x}_R^t$, the operator $V_{x_R^{t}}(\Omega_{\infty}^{(t)})$ changes under conjugation by $W({\widetilde x}_R^t,x_R^t)$. However, our boundary conditions implies that the field strength goes to zero at infinity and so the gauge potential is asymptotically flat, and consequently $W({\widetilde x}_R^t,x_R^t)$ is independent of the choice of path joining the two reference points.  Therefore, the conserved quantities are also independent of the base points. In addition, they are gauge invariant since, as shown in \cite{gauge},  $V_{x_R^{t}}(\Omega_{\infty}^{(t)})\rightarrow g_RV_{x_R^{t}}(\Omega_{\infty}^{(t)})g_R^{-1}$, with $g_R$ being  the group element, performing the gauge transformation, at $x_R^{t}$. Note in addition that if $V_{x_R^{t}}(\Omega_{\infty}^{(t)})$ has an iso-spectral evolution so does $g_cV_{x_R^{t}}(\Omega_{\infty}^{(t)})$, with $g_c \in Z(G)$. That fact has to do with the freedom we have to choose the integration constants of \rf{eqforv} and \rf{eqforv2} to lie in $Z(G)$, without spoiling the gauge covariance of \rf{basic} (see \cite{gauge}).
\begin{figure}[ht]
\centering
\subfigure[]{
   \includegraphics[width=0.039\textwidth]{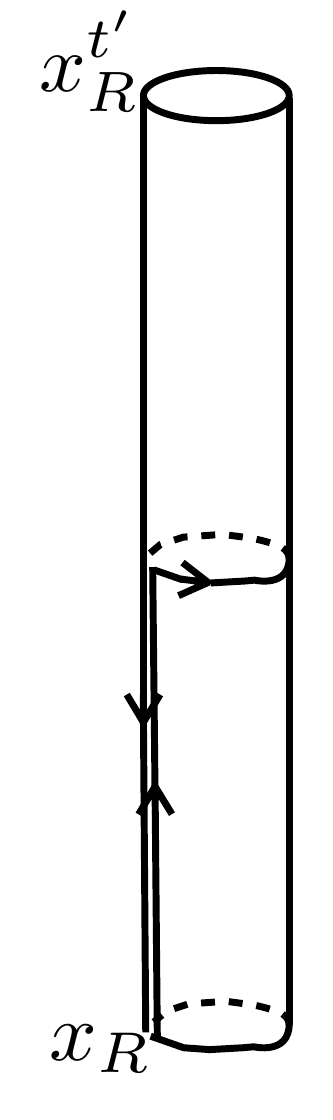}
   \label{closed_surface_a}
} \hspace{1cm}
 \subfigure[]{
   \includegraphics[width=0.12\textwidth]{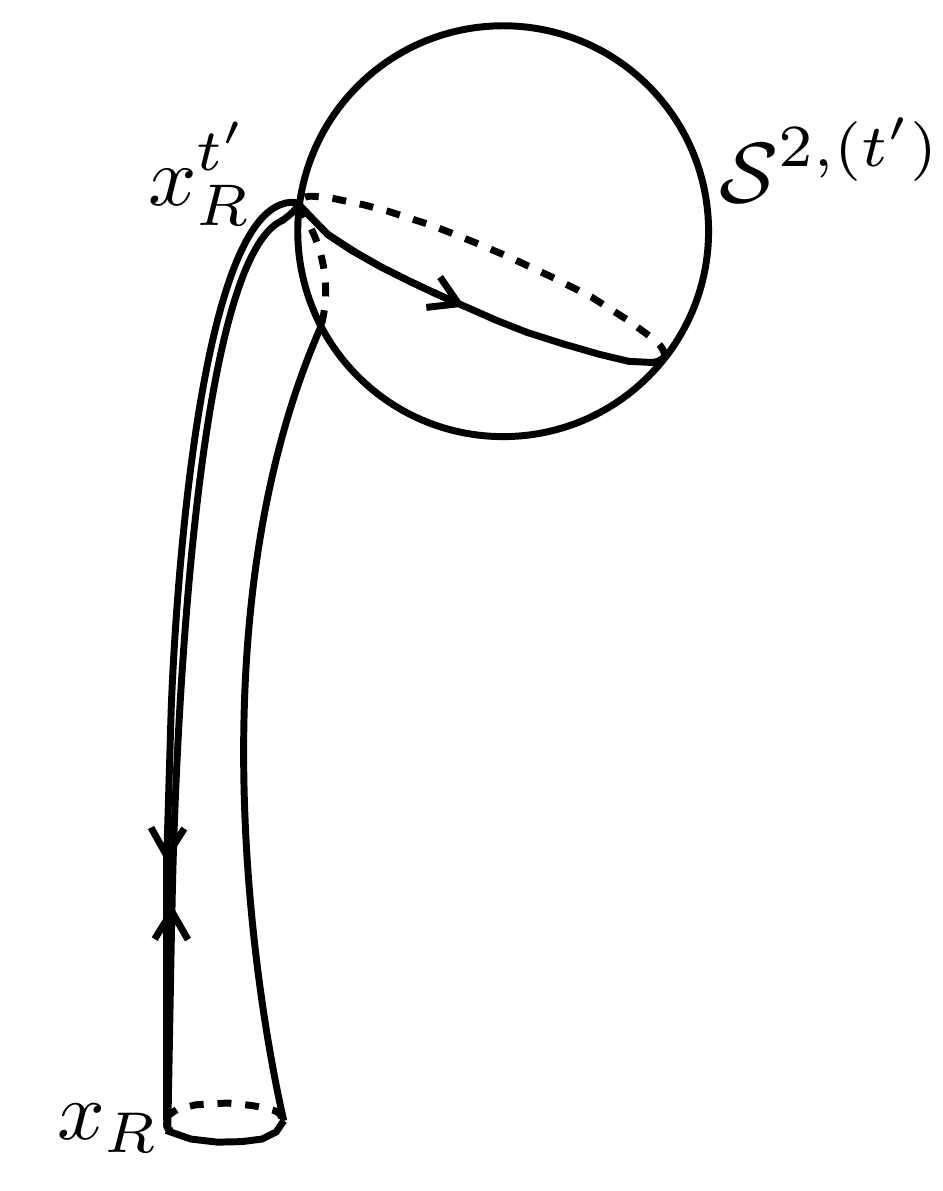}
   \label{closed_surface_b}
 } \hspace{1cm}
 \subfigure[]{
   \includegraphics[width=0.041\textwidth]{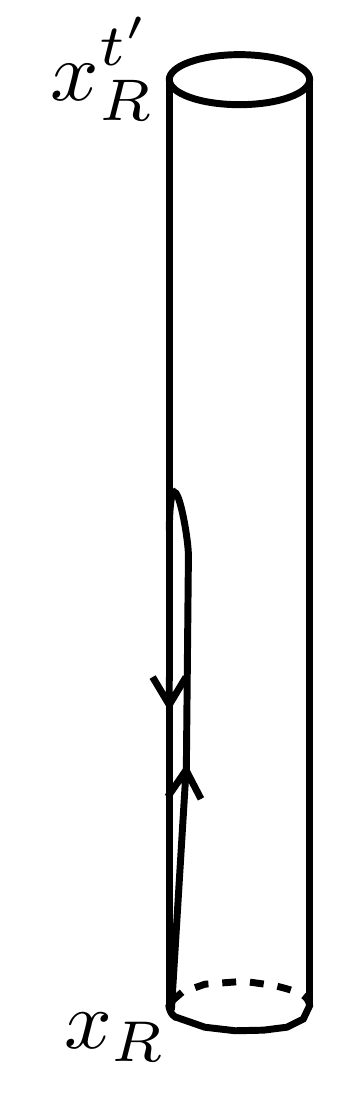}
   \label{closed_surface_c}
 }
\label{myfigure}
\caption{Surfaces of type (b) scan a hyper-cylinder $S^2\times I$.}
\end{figure}

As an example  consider a gauge theory for a gauge group $G$ spontaneously broken to a subgroup $H$ by a Higgs field $\phi$ in the adjoint representation. For a BPS dyon solution one has $E_i=\sin \theta\, D_i\phi$, and  $B_i=\cos \theta\, D_i\phi$, with $E^i=-F^{0i}$ and $B^i=-\frac{1}{2}\varepsilon^{ijk}F_{jk}$, $i,j,k=1,2,3$, with $\theta$ being an arbitrary constant angle.  At spatial infinity one has that $D^i\phi\rightarrow \frac{{\hat r}^i}{4\pi r^2}G({\hat r})$, with ${\hat r}^2=1$, $r\rightarrow \infty$, and $G({\hat r})$ being an element of the Lie algebra of $H$, which is covariantly constant, i.e. $D_{\mu}G({\hat r})=0$ \cite{olive-manton}. We have that the gauge field is asymptotically flat at spatial infinity, and so up to leading order one has $A_{\mu}=\frac{i}{e}\partial_{\mu}W\, W^{-1}$, and  so on  $S^{2,(t)}_{\infty}$ one has $G({\hat r})=WG_RW^{-1}$, with $G_R$ being the value of $G({\hat r})$ at $x_R^t$. Therefore, one has that $V_{x_R^{t}}(\Omega_{\infty}^{(t)})=P_2 e^{ie\int_{S^{2,(t)}_{\infty}}d\tau d\sigma \left[ \alpha F_{\mu\nu}^W+\beta {\widetilde F}_{\mu\nu}^W\right] \frac{dx^{\mu}}{d\sigma}\frac{dx^{\nu}}{d\tau}}=\exp\left[-ie(\alpha\,\cos \theta+\beta\,\sin\theta)G_R\right]$. Consequently, the conserved charges are given by the eigenvalues of $G_R$, which contain among them the magnetic and electric  charges of the dyon solution. Note that, even though we take $G({\hat r})$ at $x_R^t$, the eigenvalues  are independent of the choice of $x_R^t$, since   $G({\hat r})$ at different  points at infinity are related by conjugation. 

{\bf Acknowledgements} The authors are grateful to fruitful discussions with O. Alvarez, E. Castellano, P. Klimas, M.A.C. Kneipp, R. Koberle, J. S\'anchez-Guill\'en, N. Sawado and W. Zakrzewski. LAF is partially supported by CNPq, and GL is supported by a CNPq scholarship.


\begin{thebibliography}{99}
\bibitem{gauge}
  Consider a gauge transformation
$A_{\mu}\rightarrow g\, A_{\mu}\, g^{-1}+\frac{i}{e}\,\partial_{\mu}g\, g^{-1}$ 
and so $F_{\mu\nu}\rightarrow g\, F_{\mu\nu}\,g^{-1}$ and $J^{\mu}\rightarrow g\, J^{\mu}\, g^{-1}$.  From \rf{eqforw}   $W\rightarrow g_f\, W\, g_i^{-1}$, with $g_i$ and $g_f$ being the values of $g$ at the initial and final points respectively of the path  determining $W$. Consequently,  ${\cal J}$ defined in \rf{caljdef} transforms as ${\cal J}\rightarrow g_R\, {\cal J}\, g_R^{-1}$, 
with $g_R$ being the value of $g$ at $x_R$. One also has $T\(A,\tau\)\rightarrow g_R\, T\(A,\tau\)\, g_R^{-1}$, and so from  \rf{eqforv}  $V \rightarrow g_R\, V\, g_R^{-1}$.  Similarly, one sees that ${\cal K}\rightarrow g_R\,{\cal K}\,g_R^{-1}$, and so  \rf{eqforv2} also implies that $V$ transforms as $V \rightarrow g_R\, V\, g_R^{-1}$. Note however that if $V_1$ is a solution of \rf{eqforv} so is $V_2=k\,V$ with $k$ being a constant element of  $G$. Similarly, if $V_3$ satisfies \rf{eqforv2} so does $V_4=V\, h$, with $h\in G$ being constant.  Under a gauge transformation  $V_1\rightarrow g_R\,V_1\,g_R^{-1}$, and  $V_2\rightarrow g_R\,V_2\,g_R^{-1}=g_R\,k\,V_1\,g_R^{-1}$. But $k$ is any chosen constant group element and it should not depend upon the gauge field, and so it  should not change under gauge transformations. In fact, the arbitrariness associated to $k$ corresponds to the choice of integration constants in \rf{eqforv} and \rf{eqforv2}. From this point of view we should have $V_2\rightarrow k\, g_R\, V_1\,g_R^{-1}$.  The only way to establish  the compatibility is to have $k\, g_R= g_R\,k$, i.e. $k$ should be an element of the centre $Z\(G\)$ of $G$. A similar analysis applies to $V_3$ and $V_4$. Therefore, the transformation law $V \rightarrow g_R\, V\, g_R^{-1}$, and so the gauge covariance of \rf{basic}, is only valid when the integration constants in \rf{eqforv} and \rf{eqforv2} are taken in $Z\(G\)$. In such case, $V_R$ cancels out of \rf{stokes} and that is why it does not appear in \rf{basic}. Consequently \rf{basic} transforms covariantly  under gauge transformations.
\bibitem{quantize}
If for some reason at the quantum level $\alpha$ and $\beta$ are not  free parameters, then one gets  quantization conditions. Indeed, take for instance Maxwell theory, where $G=U(1)$, and  so the commutators in \rf{caljdef} drop,  the surface and volume ordering are irrelevant, and  $Q_{{\cal S}}$ is unity if  
$\int_{{\cal S}}d\zeta d\tau d\sigma{\widetilde J}_{\mu\nu\lambda}
\frac{dx^{\mu}}{d\sigma}\frac{dx^{\nu}}{d\tau}
\frac{dx^{\lambda}}{d\zeta} = \frac{2\pi n}{e\beta}$, 
with $n$  integer.
\end{thebibliography}
\end{document}